\documentclass[aps,pre,reprint,twocolum,superscriptaddress,longbibliography,showkeys,nofootinbib]{revtex4-2}

\usepackage{graphicx,xcolor}
\usepackage{amsmath,amssymb}
\usepackage{verbatim}
\usepackage{hyperref}

\begin{document}

\title{Dimensional Reduction of Solvency Contagion Dynamics on Financial Networks}


\author{Gianmarco Ricciardi}
\affiliation{Dipartimento di Fisica, Universit\`a degli Studi di Pavia, 27100, Pavia, Italy}

\author{Guido Montagna}
\affiliation{Dipartimento di Fisica, Universit\`a degli Studi di Pavia, 27100, Pavia, Italy}
\affiliation{INFN, Sezione di Pavia, 27100, Pavia, Italy}

\author{Guido Caldarelli}
\email[]{guido.caldarelli@unive.it }
\affiliation{DSMN, Universit\`a degli Studi di Venezia C\'a Foscari, 30172, Venice, Italy}
\affiliation{Institute for Complex Systems (ISC-CNR) UoS Sapienza, 00185, Rome, Italy}
\affiliation{European Centre for Living Technology, 30124, Venice, Italy}

\author{Giulio Cimini}
\affiliation{Dipartimento di Fisica, Universit\`a degli Studi di Roma Tor Vergata, 00133, Rome, Italy}
\affiliation{INFN, Sezione di Roma Tor Vergata, 00133, Rome, Italy}
\affiliation{Centro Ricerche Enrico Fermi, 00184, Rome, Italy}

\date{\today}

\begin{abstract}
Modelling systems with networks has been a powerful approach to tame the complexity of several phenomena. 
Unfortunately, such an approach is often made difficult by the large number of variables to take into consideration. 
Methods of dimensional reduction are useful tools to rescale a complex dynamical network down to a low-dimensional effective system and thus to capture the global features of the dynamics. 
Here we study the application of the degree-weighted and spectral reduction methods to an important class of dynamical processes on networks: the propagation of credit shocks within an interbank network, modelled according to the DebtRank algorithm.
In particular we introduce an effective version of the dynamics, characterised by functions with continuous derivatives that can be handled by the dimensional reduction. 
We test the reduction methods against the full dynamical system in different interbank market settings: homogeneous and heterogeneous networks generated from state-of-the-art reconstruction methods as well as networks derived from empirical e-MID data. 
Our results indicate that, for proper choices of the bank default probability, reduction methods are able to provide reliable estimates of systemic risk in the market, with the spectral reduction better handling heterogeneous networks. 
Finally we provide new physical insights on the nature and working principles of dimensional reduction methods. 
\end{abstract}


\maketitle

\section{\label{intro} Introduction}

The theory of complex networks provides a natural framework to describe the collective properties of dynamical processes taking place on large systems composed by many interacting entities \cite{Caldarelli_2007,barabasi2016network,newman2018networks}. 
The interplay between structure and dynamics in these complex systems plays a fundamental role in determining their systemic properties, such as their resilience -- namely the ability to cope with adverse events and avoid catastrophic systemic consequences \cite{LIU20221}. 
Examples applications across domains include economic and financial crises \cite{Schweitzer_2009,GUALDI201529,Bardoscia_2021}, blackouts in power grids \cite{Simpson_2016,Schafer_2018}, species mass extinctions \cite{Sole_2001,Allesina_2012,Grilli_2017} and epidemic outbreaks \cite{Pastor_2015,massaro_2018}. 
While much effort has been devoted to forecast these large-scale events \cite{Scheffer:2009wj,Scheffer_2012,Boettiger:2013uy}, no simple and universal method has yet been found because of the inherent complexity of the problem. 
Indeed for a network of $N$ nodes the evolution of the nodes' states is governed by $N$ coupled dynamical equations that depend on both the current states and the complex pattern of interactions between nodes. 
Moreover, as the number $N$ of nodes grows, the computational cost of solving $N$ coupled and often nonlinear equations increases 
and could prevent to derive manageable predictions for the system's behaviour. 
However in many case one is more interested in deriving information on some global dynamical feature of the network, 
rather than in the temporal evolution of each single node. For instance, in epidemiological studies the quantity of interest is the prevalence of the disease in the population, rather than the health state of each individual. 
In these cases a promising approach consists in using dimensional reduction methods to transform the original $N$-dimensional representation of the dynamics into a simplified version with a lower number of effective variables. 

The first attempt to apply dimensional reduction on networks has been provided by Gao {\it et al.}~\cite{Gao_2016}, who proposed a method to collapses a $N$-dimensional dynamical network into a one-dimensional equation for a global activity variable, defined as the degree-weighted average state of the network nodes. 
In this way the nodes with large degree, i.e. large number of links, contribute more to the global variable than those with a small degree. 
The underlying idea behind this approach is that the highly connected nodes have more impact on the dynamics. 
More recently, Laurence {\it et al.}~\cite{Laurence_2019} developed an independent approach that relies on the spectral property of the network under consideration --- more precisely, the dominant eigenvalues and eigenvectors of the adjacency matrix. 
The two approaches coincide in the case of uncorrelated random networks (i.e., with no degree-degree correlations), while the spectral reduction method is by definition capable to take into account the possible degree correlations. 
A thorough analysis of the two methods performed by Kundu {\it et al.}~\cite{Kundu_2021} revealed that the accuracy of the dimensional reduction strongly depends on the coefficient of variation for the equilibrium value of the relevant state variable across the nodes. This quantity in turn is determined by the dynamical system under consideration, but it is generally higher for scale-free network than for regular or Poissonian networks. 
Dimensional reduction has also been extended to the case of node-dependent the dynamical functions, though this approach works well mainly for homogeneous systems \cite{Tu_2021}. 

Both the degree-weighted and the spectral reduction have been tested on classic dynamical models in the fields of biology, ecology, epidemiology, neuroscience and population dynamics \cite{Gao_2016,Laurence_2019,Kundu_2021}, and have been recently applied to several contexts such as mutualistic ecosystems \cite{Jiang_2018}, spreading dynamics \cite{Pan_2020} and synchronization \cite{Thibeault_2020}. Surprisingly, up to now dimensional reduction has not been used in the field of economic and financial networks, despite it currently represents one of the most successful application areas of statistical physics \cite{Bardoscia_2021}. 
A paramount example is provided by the \emph{solvency contagion} dynamics, which can arise as a consequence of the bilateral exposures among financial institutions (banks, from now on). Indeed while these links allows banks to cope with liquidity fluctuations and transfer risk, they can also become channels through which distress can spread, turning an idiosyncratic shock into a systemic one. Solvency contagion played a major role in the Global Financial Crisis of 2007/08 \cite{Bardoscia2019} and thus received large attention from the literature (we remand the reader to recent reviews \citep{Huser2015,Gai2019,Jackson2021}). 
A general dynamical model for solvency contagion is represented by the DebtRank algorithm \cite{Battiston_2012,Bardoscia_2015,Barucca_2020}, which describes the following situation. When a bank suffers some losses (for instance, when one of its assets gets devalued), its creditworthiness deteriorates. As a consequence, its lenders (or counterparties) reassess the value of their claims towards the bank and thus book losses; as a consequence, their creditworthiness deteriorates, and so forth. 
As explained below, the stability of an interbank networks depends on the spectral radius of the so-called \emph{leverage} matrix \cite{Bardoscia2017}, obtained by dividing each interbank exposure by the \emph{equity} (i.e., the net value) of the creditor bank. However no simple relationship exists between the topology of the network and the spectral radius of the leverage matrix.

In this work we aim to fill the gap described above, by performing a detailed study of the application of dimensional reduction (both in its the degree-weighted \cite{Gao_2016} and spectral \cite{Laurence_2019} version) to the DebtRank dynamics. We consider different topological settings of the network, 
both homogeneous and heterogeneous graph generated from state-of-the-art statistical physics methods \cite{Cimini_2015}, 
as well as networks derived from empirical e-MID data. The aim of our work is to investigate whether dimensional reduction techniques are able to provide a sufficiently reliable description of the DebtRank dynamics. The paper is organised as follows. 
In Section~\ref{mdr} we review the two methods of dimensional reduction used in our study. 
In Section~\ref{adr} we describe the DebtRank algorithm and how to apply reduction methods in this context. 
Section~\ref{sim} describes how we generate artificial networks and run simulations of the dynamical system. 
Section~\ref{res} presents the results of our investigation, as obtained by comparing the full network simulations of DebtRank and its dimensional reductions for different network settings. Finally, in Section~\ref{conc} we draw the conclusions of our work.

\section{\label{mdr} Dimensional reduction of complex networks dynamics}

A network is defined by a set of $N$ nodes and the set of the links between them. 
We consider directed weighted networks and denote with 
$w_{ij}$ the weight of the link from node $j$ to node $i$ (with $w_{ij}=0$ if the link does not exist). 
The $N\times N$ matrix of link weights is the adjacency matrix $W$, while the (weighted) out-degree and in-degree of node $i$ are the amount of outgoing and incoming connections, respectively $k_i^{out} = \sum_j w_{ji}$ and $k_i^{in} = \sum_j w_{ij}$. 

A dynamical process on a network can be described by assigning to each node $i$ a real-valued, time dependent variable $x_i$, describing its current ``state". 
The state of each node affects the evolution of the state of its neighbours according to the underlying weighted topology of the network.
We consider dynamical processes that can be described by a set of $N$ equations (one for each node) of the form
\begin{equation}
\frac{dx_i}{dt}=F(x_i) + \sum_{j=1}^N w_{ij} \, G(x_i,x_j).
\label{dinamica}
\end{equation}
In eq.~\eqref{dinamica}, $F$ and $G$ are differentiable functions that describe the self-interaction and pairwise interaction between nodes, respectively. 
Solving such a system of $N$ coupled differential equations can be problematic. 
A slight non linearity in the interactions prevents the use of analytical approaches, and a numerical solution becomes very time consuming as $N$ increases. 
Methods of dimensional reduction can thus be used to find a small set of variables that well describe the global properties dynamical system but whose temporal evolution is easier to derive.
We now introduce the two algebraic protocols at the basis of the dimensional reduction methods proposed in the literature and explain how they allow obtaining a simplified representation of the dynamical process described by eq.~\eqref{dinamica}.

\subsection{Degree-Weighted Reduction (DWR)}

This reduction procedure is based on the simple idea that the higher the degree of a node, the higher its impact on the dynamical process \cite{Gao_2016}. The method is based on the following single global variable:
\begin{equation}
R(t) = \frac{\sum_{i=1}^N k_i^{out} x_i(t)}{\sum_{i=1}^N k_i^{out}}
\label{rdw}
\end{equation}
namely the sum of the states of the nodes, weighted by their out-degree. 
The dynamics of this variable can be enclosed into a single equation:
\begin{equation}
\dot{R} = F(R) + \alpha \, G(R,R),
\end{equation}
where the control parameter 
\begin{equation}
\alpha = \frac{\sum_{i=1}^N k_i^{out} k_i^{in}}{\sum_{i=1}^N k_i^{out}}
\label{alpha}
\end{equation}
encodes the information on the network topology.
For a complete derivation of the above equations, we refer the reader to the supplementary information of \cite{Gao_2016} 
\footnote{In \cite{Gao_2016} the quantities $R$ and $\alpha$ are denoted by $x_{\rm eff}$ and $\beta_{\rm eff}$, respectively. 
Here instead we follow the unified notation of \cite{Laurence_2019}.} 

\subsection{Spectral Reduction (SR)}
The one-dimensional version of this reduction protocol relies on the spectral properties of the adjacency matrix \cite{Laurence_2019}. 
As in the DWR method, a single reduced variable is introduced as a linear combination of the states:
\begin{equation}
R(t) = \sum_{i=1}^N a_i \, x_i(t), 
\label{rss}
\end{equation}
where $\vec{a}$ is the dominant eigenvector of $W^{\top}$, corresponding to the spectral radius $\alpha$: 
$\sum_j w_{ji} a_j = \alpha a_i$, $\forall i$. When this eigenvector is normalised, such that $\sum_i a_i=1$, the dominant eignevalue can be also expressed as
\begin{equation}
\alpha=\sum_{i=1}^N k_i^{in}\, a_i.
\label{aalpa}
\end{equation}
By defining 
\begin{equation}
\beta = \frac{\sum_{i=1}^N k_i^{in}\, a_i^2}{\alpha\sum_{i=1}^N a_i^2}
\label{bbeta}
\end{equation}
one arrives at
\begin{equation}
\dot{R}(t) = F(R) + \alpha \, G(\beta R, R).
\label{spect}
\end{equation}
In this case the network structure is encoded into two parameters: the dominant eigenvalue of the adjacency matrix $\alpha$ 
and the parameter $\beta$, which can be interpreted as a measure of the heterogeneity of the network, as shown in Appendix~\ref{app}.
For full derivation of the method we remand to \cite{Laurence_2019}, where the authors also show that for uncorrelated random networks the DWR variable $R$ of eq.~(\ref{rdw}) is an approximation of the SR variable of eq.~(\ref{rss}), while the parameter $\beta$ reduces to $1$. Therefore, the DWR formalism can be regarded as a special case of the SR procedure when applied to uncorrelated random graphs.

\section{\label{adr} Dimensional Reduction of the DebtRank dynamics}

\subsection{Definition of DebtRank}

We now discuss the applicability of the dimensional reduction method to financial contagion modelled through the DebtRank dynamics \cite{Battiston_2012,Bardoscia_2015}. 
As mentioned in the introduction, this algorithm is designed to capture the dynamics of solvency contagion within an interbank network of bilateral exposures. 
This system is represented as a weighted directed network of $N$ banks, where the generic link $i \to j$ represents the value of the interbank asset from the lender bank $i$ to the borrower bank $j$, commonly denoted as $A_{ij}$. 
For every interbank asset $A_{ij}$ in the balance sheet of bank $i$ there is a corresponding interbank liability $L_{ji} = A_{ij}$ in the balance sheet of bank $j$. 
The difference between the total assets and total liabilities (both from the interbank market and from external sources) of a bank represents its \emph{equity}, or net value. In the literature on financial contagion, the equity is a proxy of financial health: 
bank $i$ is active or solvent when its equity $E_i$ is positive, while it defaults as soon as $E_i$ vanishes 
(as it won't be able to repay its debts in full). 

Starting from interbank assets and equity values at time $t=0$, the DebtRank dynamics is triggered by exogenous shocks that cause the devaluation of the external asset and consequent decrease of equity for some banks at $t=1$. 
As a consequence, in the next time step $t=2$ the market value of the loans towards the shocked banks decreases, causing equity losses for the creditor banks, and so forth. The DebtRank assumes that assets devaluations are linear in equity losses, therefore such iterations can be described as follows:
\begin{equation}
A_{ij}(t+1) = 
\begin{cases}
A_{ij}(t)\dfrac{E_j(t)}{E_j(t-1)} &\text{ if }j\in \mathcal{A}(t-1)\\ 
A_{ij}(t)=0 & \text{ otherwise}
\end{cases}
\label{meccanismo}
\end{equation}
where $\mathcal{A}(t-1)$ is the set of active (i.e., non-defaulted) banks at $t-1$. Here the first case means that when the `wellfare' of a bank $j$ reduces, its probability of insolvency increases, and the market value of a loan $A_{ij}$ decreases proportionally. This causes an effective loss in the portfolio of the creditor bank $i$. The second case ensures that if $j$ has defaulted the value of its obligations has vanished and cannot decrease further.  

The DebtRank dynamics is framed in terms of the relative loss of equity of each bank $i$:
\begin{equation}
h_i(t)= 1-\frac{E_i(t)}{E_i(0)}.
\end{equation}
Starting from eq. \eqref{meccanismo}, the dynamical equation for these variables can be cast as (see \cite{Battiston_2012,Bardoscia_2015} for full details):
\begin{equation}
h_i(t+1) = \min\biggl[1, h_i(t) + \sum_{j=1}^N \Lambda_{ij}[h_j(t) - h_j(t-1)]\biggr]
\label{h}
\end{equation}
where $\Lambda$ is the \emph{leverage matrix}, defined as
\begin{equation}
\Lambda_{ij}=\frac{A_{ij}(0)}{E_i(0)}.
\label{leverage}
\end{equation}
We remark that the presence of the $\min(1, \cdot)$ operator in eq. \eqref{h} ensures that the equity of a bank cannot become negative. 
Indeed according to the same equation, each bank $j$ propagates shocks at time $t+1$ by mean of its last state variation: $h_j(t) - h_j(t-1)$. Hence if the bank has defaulted at $t-1$, then because of the minimum operator we have $h_j(t)=h_j(t-1)=1$ and the bank does not contribute further to equity losses, independently from the leverage matrix. 

The initial conditions of the dynamics are given, for each bank $i$, by $h_i(0)$ (equal to 0 by definition) and $h_i(1)$, representing the fractional decrease of equity due to the initial shock. Once these are set, eq. \eqref{h} can be iterated to get the equilibrium values $\{h_i^*\}_{i=1}^N$.

\subsection{Application of Dimensional Reduction Methods}\label{reduction}

In order to apply dimensional reduction techniques to the DebtRank dynamics we have to start from eq. \ref{h}, which, 
analogously to eq. \ref{dinamica}, involves a sum on the second index of the involved matrix. 
Such a common feature holds despite two different choices of notation. 
In the case of eq. \ref{dinamica}, node $i$ is influenced by another node $j$ when a direct path $j\to i$ exists, thus $w_{ij}$ represents the weight of this path. 
Equation \ref{h} instead involves paths $i \to j$ generated by an active loan $A_{ij}$ from bank $i$ to bank $j$, however financial shocks still propagate (backwards) from bank $j$ to bank $i$. 
Therefore the two formulations are consistent. 

To have eq. \eqref{h} in the form of eq. \ref{dinamica}, we use as state variables the variations of equity losses: $\Delta h(t) = h(t) - h(t-1)$. We get:
\begin{equation}
\Delta h_i(t+1) = \min \biggl[1-h_i(t), \sum_j \Lambda_{ij}\Delta h_j(t) \biggr].
\label{hhh}
\end{equation}
In order to obtain a differentiable function, we note that the minimum operator acts on a defaulted bank $i$ by selecting the null term: $1-h_i(t) \equiv 0$. We can thus introduce the probability of default of bank $i$ at time $t$, $p_i(t)$. 
As typically assumed in the DebtRank literature \cite{BardNL_2016}, we take $p_i(t)$ to be a generic monotonic function of $h_i(t)$ with extremes $p_i=0$ when $h_i=0$ and $p_i=1$ when $h_i=1$. 
Therefore we write $p_i(t)=p[h_i(t)]$ and use it as a smooth substitute for the minimum operator, approximating eq. \eqref{hhh} as:
\begin{equation}
\Delta h_i(t+1) \simeq \biggl(1-p[h_i(t)]\biggr) \sum_j \Lambda_{ij}\Delta h_j(t).
\label{happrox}
\end{equation}
As the above expression is in the form of eq. \ref{dinamica}, we can define the single variable $R(t)=\sum_i a_i h_i(t)$ representing the overall losses in the system at a given time, where the vector $\vec{a}$ will depend on the reduction method. 
The DWR is then obtained from eq. \eqref{happrox} by replacing each $h$ term with $R$ and $\Lambda_{ij}$ by $\alpha$ defined in eq. \eqref{alpha}, obtaining
\begin{equation}
\Delta R(t+1)=\bigl(1-p[R(t)]\bigr) \alpha \Delta R(t).
\label{DWiter}
\end{equation}
Instead to obtain the SR we impose that $\vec{a}$ is the dominant eigenvector of $\Lambda^{\top}$ and $\alpha$ its associated eigenvalue. We then substitute in eq. \eqref{happrox} each $h_i$ with $\beta R$, each $h_j$ with $R$ and $\Lambda_{ij}$ by $\alpha$, obtaining
\begin{equation}
\Delta R(t+1)=\bigl(1-p[\beta R(t)]\bigr) \alpha \Delta R(t).
\label{iter}
\end{equation}

\subsection{The Continuum Approximation}\label{A}

Let us assume that the time steps of the dynamics are significantly smaller than its whole duration. 
We can thus substitute the discrete variations with the time derivatives (we discuss the SR case here):
\begin{equation}
\frac{dR(t+1)}{dt} = [1-p(\beta R(t))]\alpha \frac{dR(t)}{dt}.
\end{equation}
Now we can expand the l.h.s. to the first order around time $t$:
\begin{equation}
\frac{dR(t)}{dt} + \frac{d^2R(t)}{dt^2}= [1-p(\beta R(t))]\alpha \frac{dR(t)}{dt}.
\end{equation}
Denoting by $P$ the primitive of $p$ we get
\begin{equation}
\frac{d}{dt}\left[R + \frac{dR}{dt} - \alpha\left(R-\frac{P(\beta R)}{\beta}\right)\right] = 0
\label{continuo}
\end{equation}
hence the quantity in squared brackets is a constant of the dynamics. 
We can compare its value at generic time $t$ with its initial value at $t=0$. 
Introducing simple initial conditions $R(0)=0$ and $R(1)=R_1$, and using $dR/dt |_{t=0} = R_1$ we get:
\begin{equation}
R + \frac{dR}{dt} - \alpha\left(R-\frac{P(\beta R)}{\beta}\right) = R_1+\alpha\frac{P(0)}{\beta}.
\end{equation}
It is now easy to find a closed equation for the stationary state $R^*$ of the system, by imposing the vanishing of the derivatives:
\begin{equation}
R^*(1-\alpha) +\frac{\alpha}{\beta}P(\beta R^*) = R_1 +\frac{\alpha}{\beta}P(0).
\end{equation}
To choose a proper function $P(R)$ we require that, when $R_1\to 1$ (the initial condition is full default), then also $R^*\to 1$ for any value of $\alpha$:
\begin{equation}
1-\alpha + \frac{\alpha}{\beta}P(\beta) = 1 + \frac{\alpha}{\beta}P(0).
\end{equation}
This provides the condition $P(\beta) = P(0)+\beta$ (to the same conclusion one arrives requiring $R^*\to 1$ when $\alpha \to \infty$). 
We thus can choose any primitive function that is compatible with the above condition, such as $P(R)= P(0) +\frac{1}{\beta}R^2$ and thus $p(R) = \frac{2}{\beta}R$.
The problem of this conclusion is that, for any chosen power, the parameter $\beta$ disappears from the dynamical equation.
Indeed in this case the continuum equation becomes:
\begin{equation}
R + \frac{dR}{dt} - R(1-R)\alpha = R_1
\end{equation}
which can be solved analytically, leading to
\begin{widetext}
\begin{equation}
R(t)=\frac{Q}{\alpha}\tanh\biggl[Qt + \mbox{arctanh}\frac{1}{Q}\biggl( \alpha R_1 + \frac{1-\alpha}{2}\biggr) \biggr] +\frac{\alpha - 1}{2\alpha} 
\qquad\text{ where }\qquad  Q=\biggl(\alpha R_1 + \frac{(1-\alpha)^2}{4}\biggr)^{1/2}.
\label{analitica}
\end{equation}
\end{widetext}

\section{\label{sim}Numerical Simulations}

To test the accuracy of reduction methods on the DebtRank algorithm we need two ingredients: 
1) the underlying topology of the dynamical process, that is, a weighted graph and a list of equities that represent a financial network, and 2) the full dynamical simulations as benchmark. 

\subsection{Model-Generated Networks}

Due to confidentiality constraints imposed by financial institutions, one seldom has detailed empirical information on interbank networks. 
Hence we follow the typical approach in the literature of reconstructing interbank markets from aggregate balance sheet data, namely the total interbank assets $A_i=\sum_j A_{ij}$ and liabilities $L_i=\sum_j L_{ij}$ for each bank $i$. 
Here in particular we employ the reconstruction procedure proposed in \cite{Cimini_2015}, which is grounded on statistical physics concepts applied to networks \cite{Squartini_2018,Cimini_2019}. 
The advantage of using artificially generated input data is that they allow us to explore the effectiveness of reduction methods on networks with different values of $\alpha$ and $\beta$. 

We thus start from a set of values for $\{A_i,L_i\}_{i=1}^N$, extracted from a given distribution (as explained below). 
From this input we can generate a single network instance, placing a weight on each link $i\to j$ according to a ``degree-corrected gravity model'':
\begin{equation}
w_{ij} = 
\begin{cases}
\dfrac{A_iL_j}{\Omega \cdot p_{ij}} &\text{ with probability }p_{ij}=\frac{zA_iL_j}{1+zA_iL_j}\\ 
0 & \text{ otherwise}
\end{cases}
\label{dcgm}
\end{equation}
where $z>0$ is a parameter that sets the density of the binary structure and $\Omega = \sqrt{(\sum_i A_i)(\sum_i L_i)}$ is a normalization constant. We tune $z$ for each generated network in order to obtain a link density around $10\%$ (the typical values observed in interbank markets \cite{Finger2013}). 

As described in \cite{Cimini_2015}, this procedure can be used to generate an ensemble of networks, 
where on average the reconstructed node strengths equal the input interbank assets and liabilities:
$\langle \sum_jw_{ij}\rangle = A_i$ and $\langle \sum_jw_{ji}\rangle = L_i$ $\forall i$. 
At last, to obtain values for the equities we exploit the strong correlation between the strength of a node and its equity, 
as measured from real balance sheets data \cite{Ferracci_2021}:
\begin{equation}
E_i \simeq \left[\tfrac{1}{2}(A_i+L_i)\right]^{\psi}
\label{equities}
\end{equation}
where the slope is set to $\psi=0.8$.

\subsection{Empirical Interbank Networks}

We also employ networks constructed using empirical interbank transaction data from the electronic Market of Interbank Deposits (e-MID).
As shown in \cite{Beaupain2011,Bargigli2015}, this data provide a valuable proxy for the whole structure of interbank relationships. 
Additionally it represents an unique instance of publicly available data; as such, it has been extensively analysed in the literature (we remand the reader to several papers, such as \cite{Iori2008,Fricke2015,Hatzopoulos2015,BRANDI2018}, describing the structure of the network). 

The data record every loan transaction between banks participating in the market. However, in order to have information about the underlying ``latent'' network of preferential lending relationships, data must be aggregated over a long time scale \cite{Finger2013}. 
Here we aggregate data at the yearly level, so that $w_{ij}$ represents the gross loan from bank $i$ to bank $j$ in a given year. 
After obtaining the adjacency matrix $W$, we obtain the leverage matrix using equity values derived from eq. \eqref{equities}.

\subsection{Dynamical Simulations}

Once we have a network topology, either generated or empirical as described above, we can compute the control parameter $\alpha$, 
given by eq. \eqref{alpha} for DWR and the spectral radius of the transposed leverage matrix for SR. 
In order to explore a range of $\alpha$ parameters, we follow the procedure used in \cite{Laurence_2019}: 
we start from the seed network $W_0$ (with parameter $\alpha_0$) and multiply its adjacency matrix by a global rescaling parameter $r= \alpha/\alpha_0$ to obtain a new matrix $W$, where $\alpha$ is the target control parameter. 
Note that this transformation does not affect the dominant eigenvector ($\vec{a}\equiv \vec{a}_0$ by definition) 
nor the parameter $\beta$ of the SR. Indeed from eq. \eqref{bbeta} we get:
$$\beta = \frac{\sum_i k_i^{in} a_i^2}{\alpha\sum_i a_i^2}=\frac{\sum_i r(k_0^{in})_i a_i^2}{r\alpha_0\sum_i a_i^2}= \beta_0.$$

To carry out DebtRank simulations on each network $W$, we start from a macroeconomic shock scenario for which each bank $i$ is initially healthy ($h_i(0)=0$) and then suffers from a fractional decrease of equity at $t=1$: $h_i(1) = h_1$. 
We initialise all our simulations with $h_1=0.005$, corresponding to a $0.5\%$ devaluation of all equities (a realistic value often used in the literature). We then apply eq. \eqref{h} iteratively and at each time step we compute $R(t) = \sum_i a_i h_i(t)$. 
The dynamic stops at $t^*$ when the states at time $t^*$ and $t^*+1$ are sufficiently similar. 
More precisely, we use the stop condition $|R(t^*+1) - R(t^*)|<10^{-3}$. 

In the next section we will compare simulation results with predictions from dimensional reduction methods: 
the steady states of eq. \eqref{DWiter} for DWR and eq. \eqref{iter} for SR. 
In both case we will use a highly nonlinear function 
\begin{equation}
p(h) = h^q
\end{equation}
to model the probability of bank failure. 
This choice is somehow arbitrary but as we will show it works fairly well. 
We also add to the comparison the prediction of the continuum approximation for a linear default probability, given by eq. \eqref{analitica}.

\section{\label{res}Results}

\subsection{\label{homo}Homogeneous networks}

We now report results in the case of a homogeneous system, where in the context of dimensional reduction `homogeneous' means a network setting with $\beta\simeq 1$.
This is achieved in the network generation framework described above by using a list of homogeneous total interbank assets and liabilities. In particular we can use i.i.d. binomial variables: $\forall i$,
\begin{equation}
A_i=L_i \sim B(N,\pi)
\label{Scelta}
\end{equation} 
with $N=200$ (a number of banks similar to the empirical e-MID case discussed below) and $\pi=0.1$. 

\begin{figure}[t!]
\includegraphics[width=0.5\textwidth]{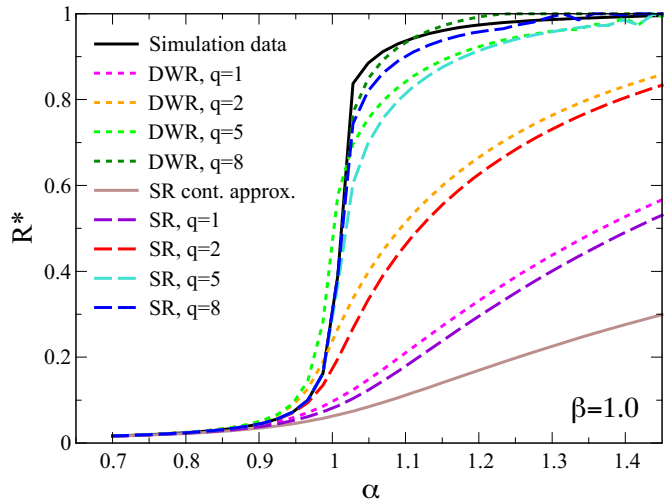}
\caption{Steady states of the reduced variable $R^*$ (the relative equity loss in the system) as a function of the spectral radius $\alpha$ of the leverage matrix, 
for a homogeneous network with $\beta \simeq 1.0$ obtained with eq. \eqref{Scelta}. 
Black solid line: full simulation of the DebtRank dynamics.
Coloured dotted lines: DWR with default probability $p(h)=h^q$.
Coloured dashed lines: SR with default probability $p(h)=h^q$.
Brown solid line: continuum approximation of eq. \eqref{analitica} with linear default probability.}
\label{cont}
\end{figure}

Figure \ref{cont} shows the stationary state of the dynamics obtained on a homogeneous network ($\beta \simeq 1.0$) for different values of the control parameter $\alpha$. 
Simulation data on the full implementation of the DebtRank dynamics features an abrupt transition for $\alpha\simeq 1$. 
Indeed, we know that the necessary condition for the convergence of eq. \eqref{h} to values $h_i^*<1$ $\forall i$ is that the spectral radius of the leverage matrix is smaller than 1; otherwise, the dynamic leads to the default of at least one bank \cite{Bardoscia_2015,Bardoscia2017}. 
In the case of a homogeneous system, banks have similar balance sheets and leverage values, and thus tend to default for similar values of $\alpha$. 
Such a steep transition is completely absent in the prediction of the continuum approximation and dimensional reduction with linear default probability. Indeed this assumption only works for values of $\alpha<1$. 
Capturing the behavior of the unstable region $\alpha>1$ instead requires a default probability that is highly non linear in the equity losses $h$. 
A good agreement with the full dynamics is in fact recovered for $q=8$. 
As expected for a homogeous system, for this choice both the DWR and SR provide similar and accurate results. 

\subsection{Heterogeneous Networks}

\begin{figure*}[t!]
     \centering
     \includegraphics[width=\textwidth]{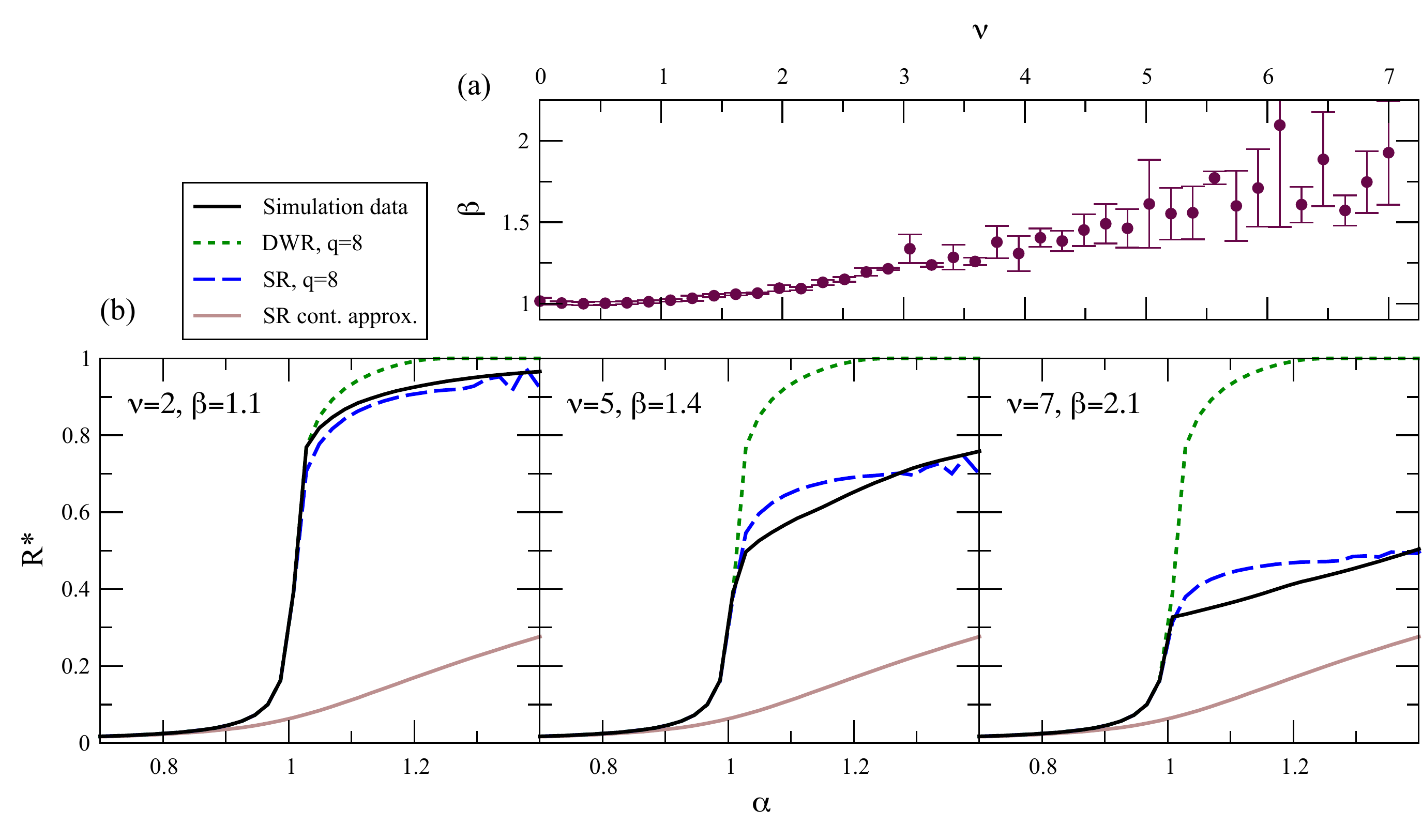}
        \caption{(a) Relation between $\nu$ and the resulting $\beta$ of a network generated by eq. \eqref{Scelta_nu}  (which does not depend on the parameter $z$ setting the link density of the network). (b) Steady states of the reduced variable $R^*$ as a function of the spectral radius $\alpha$ of the leverage matrix, for a heterogeneous network with different $\beta$ obtained with eq. \eqref{Scelta_nu} with $\nu=2$, $\nu=5$ and $\nu=7$, respectively.}
\label{ERalpha}
\end{figure*}

We then move to study more heterogeneous systems with $\beta>1$. 
This is achieved using a heterogeneous list of total interbank assets and liabilities, which can be obtained similarly to eq. \eqref{Scelta} as
\begin{equation}
A_i=L_i=x^\nu\quad\text{ where }\quad x \sim B(N,\pi)
\label{Scelta_nu}
\end{equation} 
i.e. as powers of binomial variable, $\forall i$ (we again use $N=200$ and $\pi=0.1$). 
As shown in Figure \ref{ERalpha}(a), by changing the exponent $\nu$ it is possible to increase the heterogeneity of the network, in terms of the coefficient $\beta$ of the output leverage matrix. 
Figure \ref{ERalpha}(b) shows the stationary state of the dynamics, as a function of $\alpha$, for heterogeneous network obtained using different values of $\nu$. Notably, also for high values of $\beta$ the SR approach remains accurate for a wide range of $\alpha$ values, in particular around the transition at $\alpha=1$, while the DWR behaviour is independent of $\beta$ and thus leads to inaccurate results for $\beta>1$. 

\begin{figure*}[t!]
     \centering
     \includegraphics[width=\textwidth]{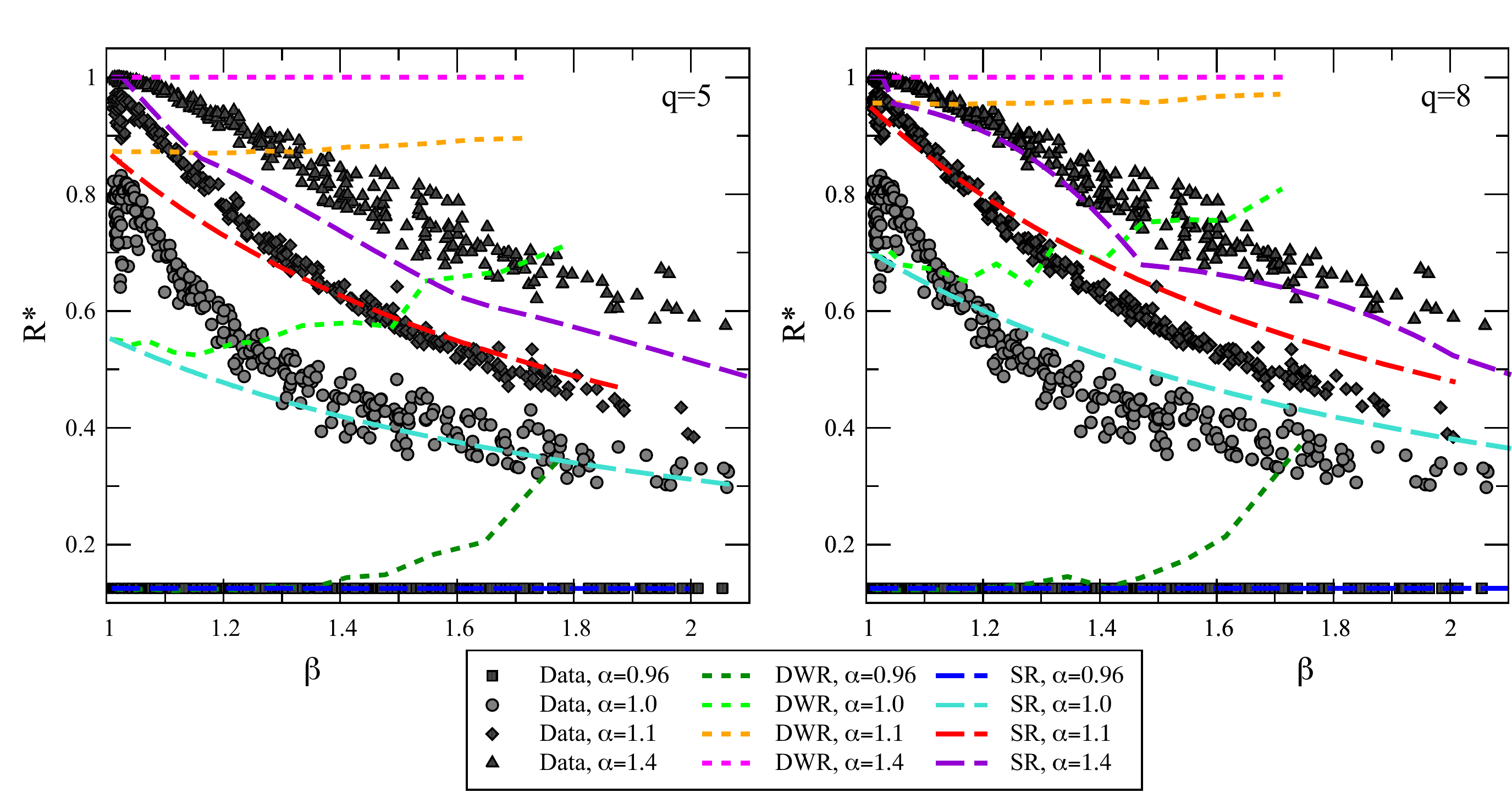}
         \caption{Steady states of the reduced variable $R^*$ as a function of the heterogeneity $\beta$ of the network, for different values of the control parameter $\alpha$. Networks have been generated using eq. \eqref{Scelta_nu}, tuning the exponent $\nu\in[0,10]$. 
Cloud of points: results of the full DebtRank dynamics.
Dashed and dotted lines: solutions of the SR and DWR, respectively, with $p(h)=h^q$ with $q=5$ and $q=8$ (left and right panel, respectively). 
Note that the choice $q=8$ leads in general to a better agreement between SR and simulation results for small values of $\beta$, where the method is well defined, whereas smaller powers like $q=5$ are more accurate for larger $\beta$.}
         \label{TreAlpha}
\end{figure*}

The same picture is obtained by plotting the steady states of the reduced variable $R$ as a function of the heterogeneity parameter $\beta$, for different values of the control parameter $\alpha$. 
As Figure \ref{TreAlpha} shows, the SR solutions with nonlinear default probabilities are in good agreements with the cloud of points, corresponding to full DebtRank simulations for an ensemble of networks with various $\beta$. 
On the contrary, the accuracy of DWR predictions decays with increasing $\beta$, since the method does not account for the network heterogeneity.

Note how, for high values of $\alpha$, the lines corresponding to the steady state of the SR 
as a function of $\alpha$ (Fig. \ref{ERalpha}) or as a function of $\beta$ (Fig. \ref{TreAlpha}) become irregular. 
This is due to the iterative solution of the reduced equation \eqref{iter}, which is well defined only until $R\le \beta^{-1}$, otherwise the term $1-p(\beta R)$ becomes negative -- while in the DebtRank dynamics the total amount of stress can only increase. 
Hence when the iterations would reach $R>\beta^{-1}$ we effectively stop them; however the stopping time $t^*$ decreases with $\alpha$, as the latter represents the amount of the increment in $R$ at each time step. Indeed if for a specific $\alpha$ we have at $t^*$ that $R\equiv \beta^{-1}$, if we increase $\alpha$ by a tiny amount then the stopping time will decrease by one, where $R$ will be smaller than the previous value. 
Increasing $\alpha$ further will lead to an increase of $R$ until $t^*$ decreases again and $R$ drops, and so on.

We also tried an alternative method to generate even more heterogeneous networks: drawing total interbank assets and liabilities directly from a power-law distribution, with exponent $-3\nu$. Results, reported in Appendix \ref{app2}, confirm the outcome of the previous analysis: 
the SR method can well approximate the dynamics on heterogeneous networks, contrarily to the DWR that works well only for low values of $\beta$. 

\subsection{e-MID data}

\begin{figure*}[t!]
     \centering
     \includegraphics[width=\textwidth]{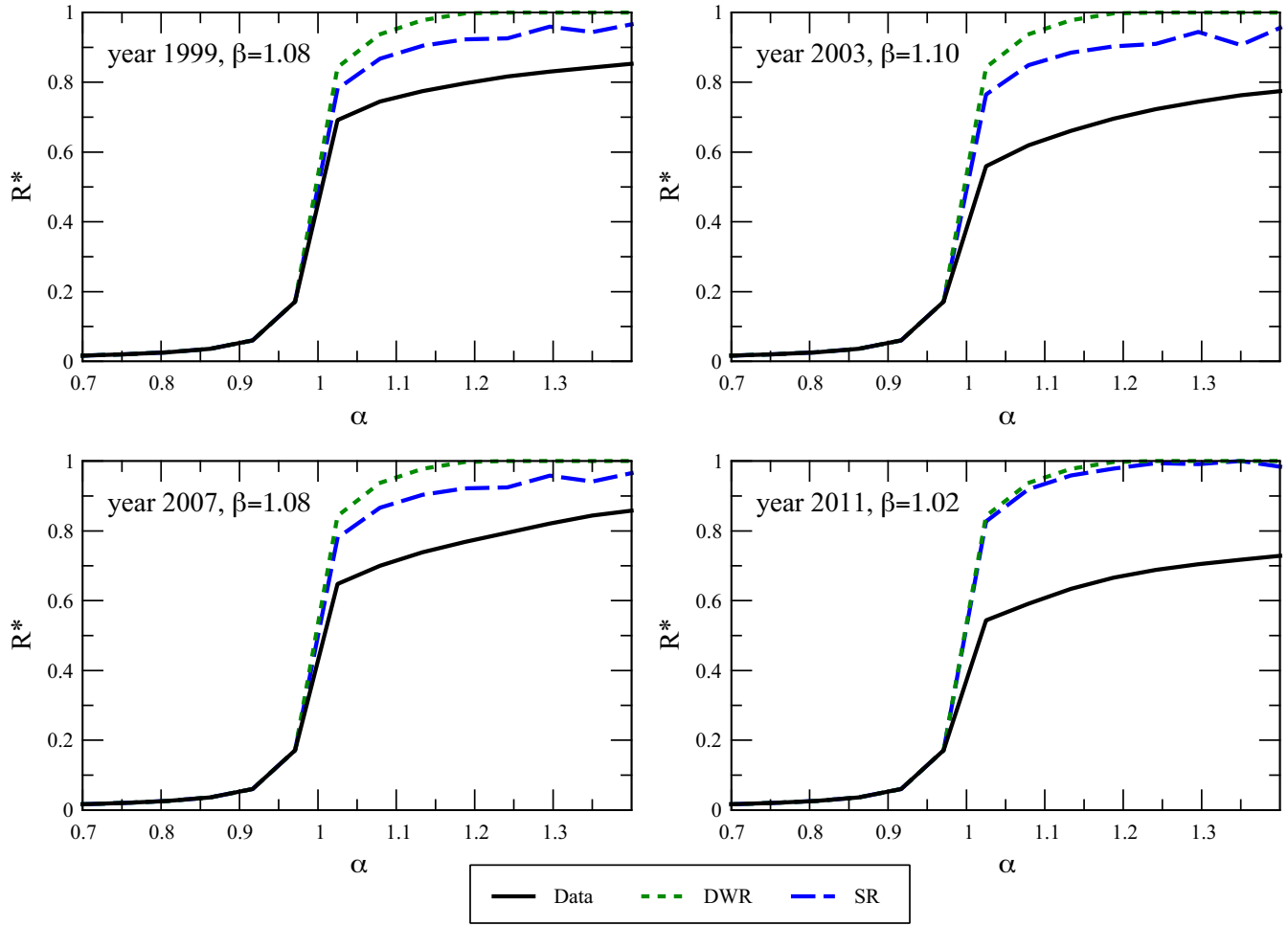}
\caption{Steady states of the reduced variable $R^*$ as a function of the spectral radius of the leverage matrix, for the e-MID network in various representative years.}
\label{emid}
\end{figure*}

Finally we consider empirical networks from e-MID transaction data. 
As shown in Figure \ref{emid}, these networks are characterised by small heterogeneity values ($\beta\simeq 1$), 
so that both DWR and SR are able to properly capture the transition of the systemic risk variable, with SR systematically performing better. 
However, after the transition the real simulations do not converge to $R=1$, corresponding to full default. 
This is due to the presence in the data of some bank with zero out-degree (i.e., no lending), which by definition cannot suffer losses and go bankrupt. These banks amount to 1\% of the total in 1999, a percentage that grows to 2\% in 2003 and 2007 and reaches 8\% in 2011, where the number of banks is also halved (124 in 2011 versus 215 in 1999). The size of these banks determines the maximal amount of relative equity loss in the system; however spectral reduction methods cannot take this aspect into account. 

\section{\label{conc} Conclusions}

In this work we studied how spectral reduction techniques -- both the degree-weighted reduction by \cite{Gao_2016} and the spectral reduction by \cite{Laurence_2019} -- can be applied to the DebtRank dynamics \cite{Battiston_2012,Bardoscia_2015}, 
which models a solvency contagion process on an interbank networks. 
We introduced an effective differentiable version of the dynamics that can be handled by reduction methods, 
and tested the derived reduced equation on homogeneous and heterogeneous financial networks that we generated by state-of-the-art  reconstruction procedures \cite{Cimini_2015}, as well as on empirical e-MID networks. 
We found that the spectral reduction systematically outperforms the degree-weighted one thanks to the presence of an additional parameter, $\beta$, which relates to the heterogeneity of the network. Indeed $\beta$ is the main parameter that affects the accuracy of the spectral reduction, allowing to obtain a remarkable agreement between the behavior of the reduced variable and that of the full dynamical system. 

Our results have both theoretical and practical implications. 
Firstly, we showed that reduction techniques can be successfully applied to dynamical systems that are more general than those described by eq. \eqref{dinamica}. Improvements in this direction would consist in using default probability functions that are not arbitrary but derived from more principled financial arguments. 
Secondly, for the considered case of a delocalised initial shock, we confirmed that the total equity losses in a financial system monotonically increase with the spectral radius of the leverage matrix \cite{Bardoscia_2015,Bardoscia2017}. More importantly, we showed that these losses are higher for more homogeneous systems, contributing to the growing literature on network sensitivity of systemic risk (see the discussion in \cite{Ramadiah2020,Ferracci_2021}). Further analysis of localized initial shocks represents an interesting avenue for future research. 

Finally we remark that the higher performance of spectral reduction with respect to the degree-weighted counterpart can be expected, given that the former method represents a network with two parameters while the latter only uses one. 
Therefore we believe that future efforts in the foundational theory of reduction techniques should be aimed at finding general procedures to encode networks in a richer parameter set to be used within the reduced equations.


\appendix

\section{\label{app}Topological Interpretations of the $\alpha$ and $\beta$ Parameters of the Spectral Reduction}

The power method allows to efficiently compute the dominant eigenvector of a matrix, 
by repeated applications of the matrix itself to an arbitrary starting vector (which must not be orthogonal to the eigenvector). 
Consider for simplicity an undirected and unweighted graph, and let $W$ be its (symmetric) adjacency matrix. 
If $W$ is primitive, by the Perron Theorem we know that its dominant eigenvalue $\alpha$ is a positive real number, with the corresponding eigenvector having only positive components. We can thus use the power method with a starting vector of ones, 
so that at the $n$-th iteration the element $(W^n)_{ij}$ gives the number of possible paths of length $n$ starting from $i$ and ending in $j$, 
while $\mu_i(n)=\sum_j (W^n)_{ij}$ is the total number of paths of length $n$ starting from node $i$.
Consequently, the (normalized) dominant eigenvector of $W$ satisfies:
\begin{equation}
a_i = \lim_{n\to \infty} \frac{\mu_i(n)}{\sum_j \mu_j(n)} \qquad \forall  i.
\label{autvec}
\end{equation}
Hence the dominant eigenvector (and thus the SR method) values a node proportionally to the number of infinitely long paths starting from that node. On the other hand, by using degree weights the DWR takes into account only paths of length 1. 
Therefore, SR and DWR can be seen as laying at the extremes of the power method:
\begin{gather}
\vec{v}^{(0)}=\vec{1} \xrightarrow{W}
\underbrace{\vec{v}^{(1)}=\vec{k}}_{\text{DWR}} \xrightarrow{A}
\vec{v}^{(2)} \xrightarrow{W}\dots \xrightarrow{A}
\underbrace{\vec{v}^{(\infty)}=\vec{a}}_{\text{SR}}.
\end{gather} 
Note that the degree only represents a local centrality measure, failing to provide information on which nodes are actually connected. 
In contrast, the dominant eigenvector yields a more refined centrality metric since it contains the information on how each node is connected with the rest of the network \cite{Laurence_2019}. In particular, using eq. \eqref{autvec} the dominant eigenvalue can be expressed as:
\begin{equation}
\alpha = \lim_{n\to \infty} \frac{\sum_i\mu_i(n+1)}{\sum_i \mu_i(n)}.
\label{autdom}
\end{equation}
Thus, $\alpha$ is the ratio between paths of length $N+1$ and $N$ in the graph.

\begin{figure}[t!]
     \centering
          \includegraphics[width=0.45\textwidth]{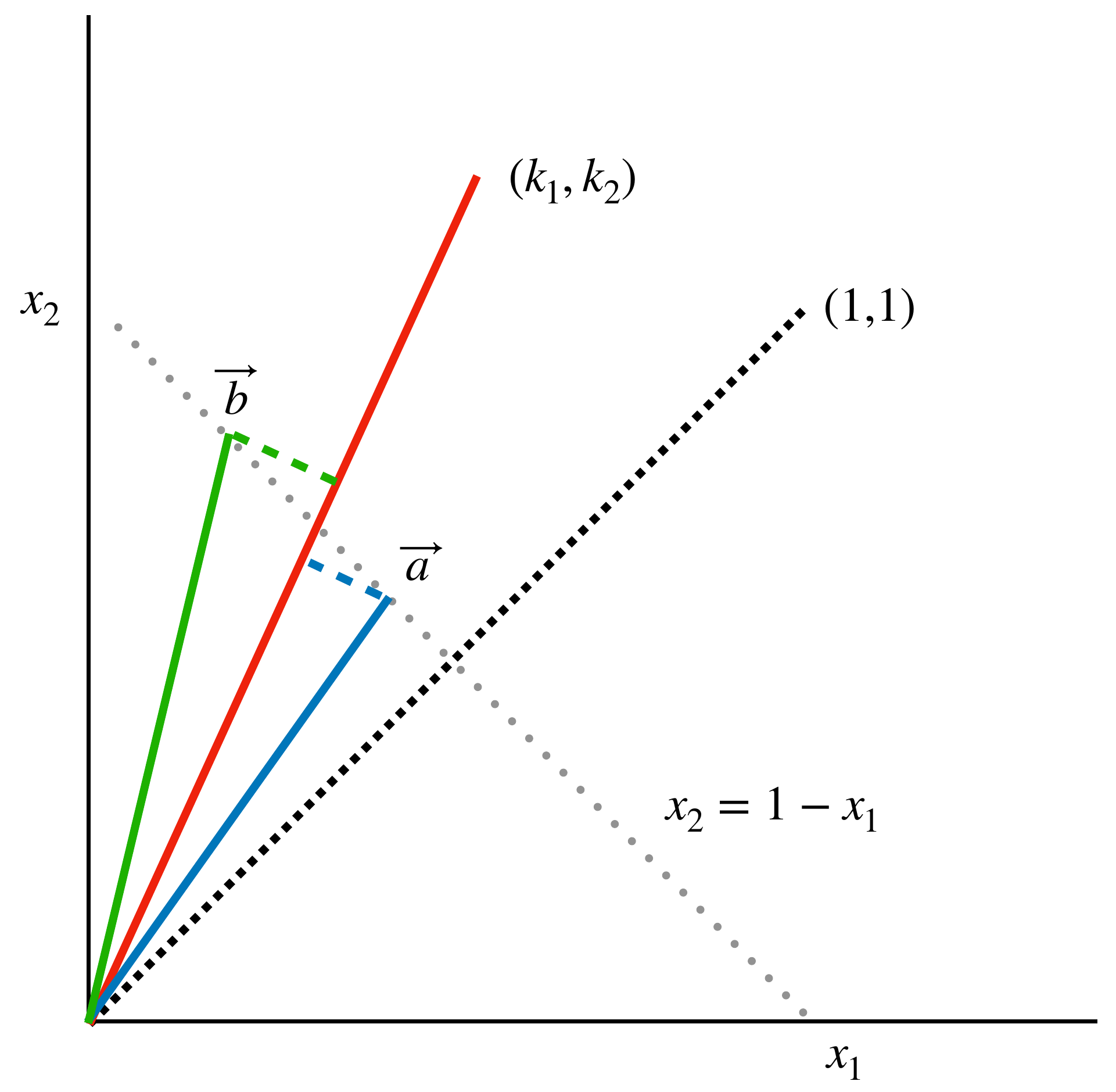}
        \caption{Geometrical construction to understand the meaning of the $\beta$ parameter.}
         \label{fig.beta}
     \end{figure}

On the other hand, the parameter $\beta$ defined in equation \eqref{bbeta} can be interpreted as a measure of the network's heterogeneity. 
First, by defining $b_i= a^2_i/(\sum_j a^2_j)\;\forall i$ we can combine eqs. \eqref{aalpa} and \eqref{bbeta} to rewrite $\beta$ as 
\begin{equation}
\beta = \frac{\sum_{i=1}^N b_i k_i^{in}}{\sum_{i=1}^N a_i k_i^{in}}
\label{betaapp}
\end{equation}
Consider again for simplicity an undirected and unweighted graph, and take any two nodes that we label as 1 and 2. 
Figure \ref{fig.beta} depicts a two-dimensional space where we can represent relevant vectors associated to these two nodes.
In particular we can represent the vector of degrees, $\vec{k}=(k_1, k_2)$, the bisector $(1,1)$, as well as the components of the dominant eigenvector $\vec{a}$ and of the vector $\vec{b}$ defined above. 
If these latter two vectors are normalized (so that $a_1+a_2=b_1+b_2=1$), graphically their heads lie on the line $x_2 = 1 - x_1$. 
We further note that when $a_1<a_2$ we have
\begin{equation}
\frac{b_1}{b_2} = \frac{a_1^2}{a_2^2} < \frac{a_1}{a_2} < 1.
\end{equation}
In this case, walking on the line $x_2 = 1 - x_1$, vector $\vec{a}$ lies between $\vec{b}$ and the bisector. 
Since eq. \eqref{betaapp} says that $\beta$ is the projection of the vector $\vec{b}$ on $\vec{k}$ over the projection of $\vec{a}$ on $\vec{k}$, we have the following three cases.
\begin{itemize}
\item To have $\beta =1$, the two projections should be equal, which can happen only if $\vec{k}$ is parallel to the bisector: the two nodes have the same degree and the network is homogeneous.
\item We have $\beta >1$ as soon as $\vec{k}$ is on the same side of $\vec{a}$ and $\vec{b}$ with respect to the bisector, which means $k_1<k_2$. The larger this difference (that is, the more the network is heterogeneous), the higher the value of $\beta$. 
\item Finally, to have $\beta<1$ we would need $k_1>k_2$, which is an uncommon and peculiar situation for which node degree is not representative of its eigenvector centrality.
\end{itemize}

\section{\label{app2}Additional Results on Heterogeneous Networks}

\begin{figure*}[t!]
     \centering
     \includegraphics[width=\textwidth]{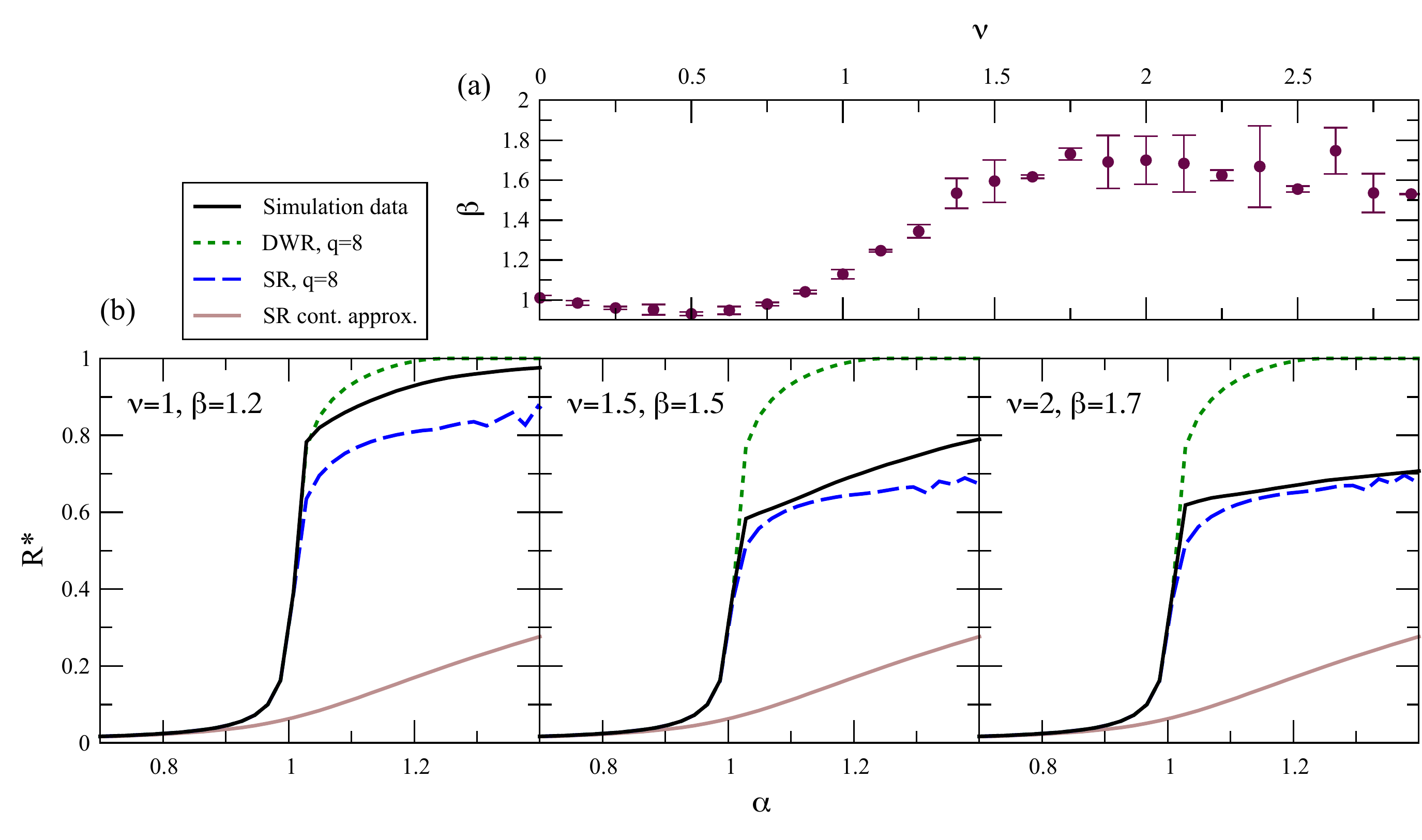}
        \caption{(a) Relation between $\nu$ and the resulting $\beta$ of a network generated by eq. \eqref{Scelta_3nu}  (which does not depend on the link density). (b) Steady states of the reduced variable $R^*$ as a function of the spectral radius $\alpha$ of the leverage matrix, for a heterogeneous network with different $\beta$ obtained with eq. \eqref{Scelta_3nu} with $\nu=1$, $\nu=1.5$ and $\nu=2$, respectively.}
\label{BAalpha}
\end{figure*}

\begin{figure}[t!]
\centering
         \includegraphics[width=0.5\textwidth]{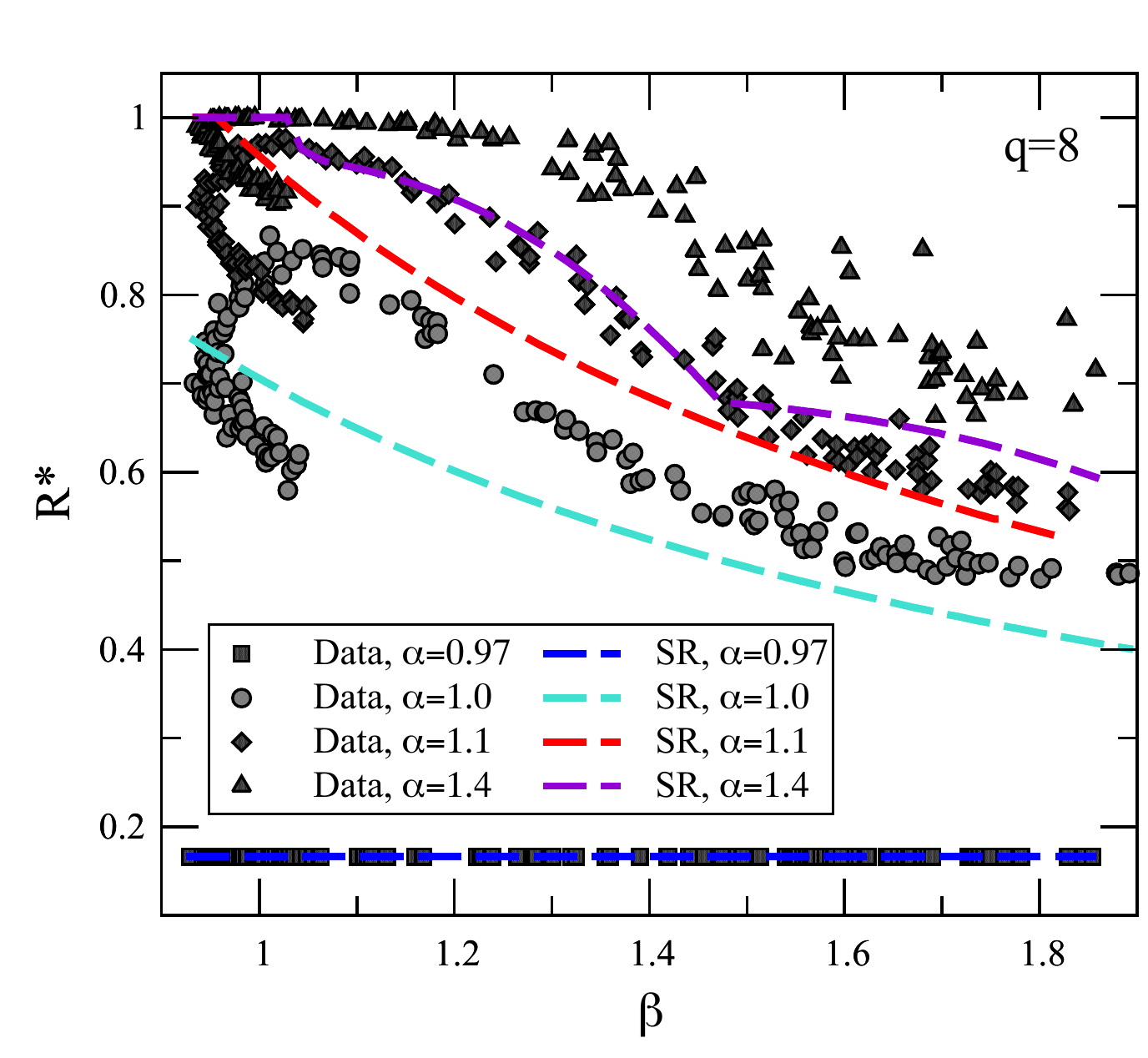}
          \caption{Steady states of the reduced variable $R^*$ as a function of the heterogeneity $\beta$ of the network, for different values of the control parameter $\alpha$. Networks have been generated using eq. \eqref{Scelta_3nu}, tuning the exponent $\nu\in[0,2.5]$. 
Cloud of points: results of the full DebtRank dynamics.
Dashed and dotted lines: solutions of the SR and DWR, respectively, with $p(h)=h^8$.}
         \label{TreAlpha2}
\end{figure}

Here we show the results of the dimensional reduction when the network topology is generated using total interbank assets and liabilities that are obtained as powers of variables drawn from a power-law distribution: $\forall i$,
\begin{equation}
A_i=L_i=y^\nu\quad \text{ where }\quad P(y) \sim y^{-3}
\label{Scelta_3nu}
\end{equation} 
again using a network of size $N=200$ and $y_{min}=3$. Also in this case, as shown in Figure \ref{BAalpha}(a), by changing the exponent $\nu$ we can increase the heterogeneity of the network in terms of the parameter $\beta$ of the output leverage matrix. 
Figures \ref{BAalpha}(b) and \ref{TreAlpha2} show the stationary states of the dynamics, respectively as a function of $\alpha$ and $\beta$, for heterogeneous network obtained using the above input with different values of $\nu$. Notably, also for high values of $\beta$ the SR approach remains accurate for a wide range of $\alpha$ values, in particular around the transition at $\alpha=1$, while this is not the case for DWR. 
However, in the lower range of $\beta$ values a bifurcation of the stationary states occurs. This means that a third structural parameter may be necessary to properly reduce the dynamics.

%

\end{document}